# Simultaneous Realization of Coherent Perfect Absorber and Laser by Zero-Index Media with both Gain and Loss

Ping Bai[1,2], Kun Ding[3], Gang Wang[1], Jie Luo[1], Zhao-Qing Zhang[3], C.T. Chan[3], Ying Wu[2]\*, Yun Lai[1]\*

[1]*College of Physics, Optoelectronics and Energy & Collaborative Innovation Center of Suzhou Nano Science and Technology, Soochow University, Suzhou 215006, China*
[2]*Division of Computer, Electrical and Mathematical Sciences and Engineering, King Abdullah University of Science and Technology (KAUST), Thuwal 23955-6900, Saudi Arabia*
[3]*Department of Physics and Institute for Advanced Study, The Hong Kong University of Science and Technology, Clear Water Bay, Kowloon, Hong Kong, China*

Email: laiyun@suda.edu.cn
　　　Ying.Wu@kaust.edu.sa

We investigate a unique type of zero-index medium with both gain and loss (ZIM-GL), whose effective permittivity and permeability are purely imaginary and of opposite signs. We show analytically that, by using a slab of ZIM-GL with equal magnitude of loss and gain, coherent perfect absorber (CPA) and laser, i.e. the so-called "CPA-laser", can be achieved simultaneously. CPA-laser has been found previously in parity-time (PT) symmetric systems. However, the underlying physics in a PT-symmetric system is distinct from that in a ZIM-GL. By designing a photonic crystal (PC) composed of core-shell rods, with loss and gain distributed in either the core or the shell, we have realized such a ZIM-GL. The CPA-laser functionality of such a PC is also confirmed in our numerical simulations. Our work provides a different approach for simultaneous realization of CPA and laser besides PT-symmetric systems.

# I. INTRODUCTION

Very recently, parity-time (PT) symmetric systems have attracted intensive research interests. In quantum mechanics, this newfangled system is described by a non-Hermitian Hamiltonian with complex potential of $V(x)=V^*(-x)$, which shares the same eigenstates with the PT operator, and thus can exhibit real eigenvalue spectra. In optics, the PT-symmetric system is characterized by complex permittivity exhibiting a symmetric spatial profile of $\varepsilon(x)=\varepsilon^*(-x)$ [1-4], i.e. with gain and loss symmetrically distributed in space. Interestingly, the gain in such optical PT-symmetric systems plays a far more complicated role than simply loss compensation. Numerous intriguing optical phenomena have been observed in optical PT-symmetric systems, such as PT-symmetry phase transition and exceptional points [5-8], unidirectional invisibility [9,10], etc. Especially, an optical PT symmetric medium that can simultaneously behave as a coherent perfect absorber with 100% absorption [11-14], and as a laser oscillator [15,16] by simply adjusting the amplitudes and phases of incoming waves has been proposed, which is denoted as the "CPA-laser" [17-18].

In this paper, instead of studying spatially distributed loss and gain, as in PT symmetric systems, we investigate the case where loss and gain are both uniformly distributed in space but appear in different parameters of a homogeneous medium [32-34]. More specifically, we consider a unique zero-index medium characterized by purely imaginary relative permittivity and relative permeability, i.e., $\varepsilon = \pm|\varepsilon|i$ and $\mu = \mp|\mu|i$, delineating the loss (for positive sign) and gain (for negative sign), respectively. It is interesting to note the fact that the refractive index $n=\sqrt{|\varepsilon||\mu|}$ is a real number implies the existence of plane wave solutions for both electric and magnetic fields. However, as will be demonstrated later, such a medium does not transport energy because plane waves of electric field and magnetic field always differ by a phase of $\pm\pi/2$, leading to a zero time-averaged Poynting vector. We show explicitly that when a slab of such a medium is embedded in air, CPA and laser can occur due to an uneven spatial distributions of electric and magnetic fields caused by the boundary condition.

Since the real parts of both $\varepsilon$ and $\mu$ are zero [19-23], we call such a unique type of electromagnetic medium as a zero-index medium with both gain and loss (ZIM-GL). The CPA-laser is achieved, when $|\varepsilon|=|\mu|$. Since the ZIM-GL considered here is homogeneous and possesses no PT-symmetry. Therefore, our work demonstrates that PT symmetry is not the necessary requirement for the realization of CPA-laser.

To demonstrate the feasibility of our proposal, we design a photonic structure which behaves effectively as ZIM-GL. The structure is composed of core-shell dielectric rods, with loss and gain distributed in the permittivity of the cores and shells, respectively. Interestingly, the effective medium of such a structure exhibit a positive purely imaginary permittivity and a negative purely imaginary permeability of the same magnitudes, i.e. ZIM-GL. Furthermore, we have numerically demonstrated its functionality as a CPA-laser. We have observed the switching between the CPA state with 100% absorption and the lasing state of stimulated emission, by simply tuning the symmetry of incident waves. These results show that realization of CPA-laser based on ZIM-GL is a feasible and promising approach.

## II. THEORETICAL ANALYSIS ON CPA AND LASER BY ZIM-GL

We start our theoretical analysis by considering two counter-propagating plane waves of frequency incident onto a slab of ZIM-GL. The schematic graph is shown in Fig.1 (a). For simplicity, we have chosen our system to be symmetric about $x=0$. The electric fields in the background (air) and slab can be written as

$$\begin{cases} E_1 = a_1 e^{ik_0 x} + b_1 e^{-ik_0 x} & (x \leq -d/2) \\ E_2 = a_2 e^{-ik_0 x} + b_2 e^{ik_0 x} & (x \geq d/2) \\ E_3 = c e^{ikx} + d e^{-ikx} & (-d/2 \leq x \leq d/2) \end{cases}, \quad (1)$$

where $k_0 = \omega/c_0$, $k = k_0 n$. $a_i$, $b_i$, $c$ and $d$ are the coefficients of the right and left propagating waves. By matching the standard boundary conditions, it is easy to obtain the following transfer matrix:

$$\begin{pmatrix} b_2 \\ a_2 \end{pmatrix} = \mathbf{M} \begin{pmatrix} a_1 \\ b_1 \end{pmatrix}, \tag{2}$$

with $\mathbf{M} = \begin{pmatrix} M_{11} & M_{12} \\ M_{21} & M_{22} \end{pmatrix}$ and

$$M_{11} = e^{-idk_0}\left(\cos(\eta) + i\frac{n^2 + \mu^2}{2n\mu}\sin(\eta)\right), \tag{3}$$

$$M_{22} = e^{idk_0}\left(\cos(\eta) - i\frac{n^2 + \mu^2}{2n\mu}\sin(\eta)\right), \tag{4}$$

$$M_{12} = i\frac{n^2 - \mu^2}{2n\mu}\sin(\eta), \tag{5}$$

$$M_{21} = -M_{12}, \tag{6}$$

where $\eta = nk_0 d$. It is straightforward to show from Eqs. (3-6) that det(**M**)=1.

In terms of scattering matrix, **S**, Eq. (2) can be rewritten as

$$\begin{pmatrix} b_2 \\ b_1 \end{pmatrix} = \mathbf{S}\begin{pmatrix} a_1 \\ a_2 \end{pmatrix} = \begin{pmatrix} t_1 & r_2 \\ r_1 & t_2 \end{pmatrix}\begin{pmatrix} a_1 \\ a_2 \end{pmatrix}, \tag{7}$$

where $t_1$, $t_2$ and $r_1$, $r_2$ represent the transmission and reflection coefficients of the slab for a plane wave normally incident from the left and right, respectively. By using the condition det(**M**)=1, we find

$$t_1 = t_2 = t = 1/M_{22} \text{ and } r_1 = r_2 = r = M_{12}/M_{22}. \tag{8}$$

The equalities of $t_1 = t_2$ and $r_1 = r_2$ are expected due to reciprocity and parity symmetry of our system, respectively. In the presence of gain and loss, the system does not have time-reversal symmetry. Here, the **S**-matrix has the following simple form:

$$\mathbf{S} = \begin{pmatrix} t & r \\ r & t \end{pmatrix} = \frac{1}{M_{22}}\begin{pmatrix} 1 & M_{12} \\ M_{12} & 1 \end{pmatrix}, \tag{9}$$

with eigenvalues $\lambda_\pm = t \pm r = (1 \pm M_{12})/M_{22}$ and eigenstates $\varphi_\pm^T = (1, \pm 1)$. For any two incident beams, the two outgoing beams can obtained from Eqs. (7) and (9) as

$$\begin{pmatrix} b_2 \\ b_1 \end{pmatrix} = \frac{1}{2}[(a_1 + a_2)(t+r)\varphi_+ + (a_1 - a_2)(t-r)\varphi_-]. \tag{10}$$

In order to obtain CPA, the right-hand side of Eq. (10) has to vanish. This can be achieved either by setting $a_1 = a_2$ (in-phase excitation) and find the condition for $t + r = 0$, or by setting $a_1 = -a_2$ (out-of-phase excitation) and find the condition for $t - r = 0$. While for lasers, the conditions can be determined from the divergence condition of $t+r$ or $t-r$.

Next, we shall focus on the special case of ZIM-GL with purely imaginary permittivity and permeability of opposite signs. Such an ZIM-GL can be represented by effective parameters of $\varepsilon = i\alpha$ and $\mu = -i\beta$ with $\alpha > 0, \beta > 0$ or $\alpha < 0, \beta < 0$. Since the case of $\alpha < 0, \beta < 0$ only represents a switch of gain and loss (as well as CPA and lasing modes), we will only consider the case of $\alpha > 0, \beta > 0$ in the discussions below.

When $\alpha > 0, \beta > 0$, the condition of $1 + M_{12} = 0$ is

$$\frac{\alpha + \beta}{2\sqrt{\alpha\beta}} \sin(\eta) = 1, \eta = \sqrt{\alpha\beta} k_0 d . \tag{11}$$

Since $\alpha + \beta > 2\sqrt{\alpha\beta}$, Eq. (11) has two solutions within the region of $0 < \eta < \pi$. When $\alpha > \beta > 0$, the solution of Eq. (11) with $0 < \eta < \pi/2$ satisfies $M_{22} = \cos(\eta) + \frac{\alpha - \beta}{2\sqrt{\alpha\beta}} \sin(\eta) \neq 0$. Therefore, we obtain a CPA mode for the in-phase excitation $(a_1 = a_2)$, i.e. $t + r = (1 + M_{12})/M_{22} = 0$. The other solution of Eq. (11) appearing in the region of $\frac{\pi}{2} < \eta < \pi$ is actually not a CPA solution when $\alpha \neq \beta$. because at this solution we also find

$$M_{22} = \cos(\eta) + \frac{\alpha - \beta}{2\sqrt{\alpha\beta}} \sin(\eta) = 0 , \tag{12}$$

and the ratio of 1+ and is non zero, i.e.,

$$t + r = \lim \frac{1+M_{12}}{M_{22}} = e^{-ik_0 d} \frac{\alpha^2 - \beta^2}{(\alpha - \beta)^2 - 4\alpha\beta} . \tag{13}$$

It is interesting to point out that the second solution with $M_{22} = 0$ and $1 - M_{12} \neq 0$ actually corresponds to the laser mode with two divergent out-of-phase output beams ($b_1 = -b_2$) as long as the input beams are not in-phase ($a_1 \neq a_2$). This can be seen from Eq. (10) with $t - r = (1 - M_{12})/M_{22} = \infty$. In short, the condition of $1 + M_{12} = 0$ gives one CPA mode for two in-phase input beams ($a_1 = a_2$) and one lasing mode with two out-of-phase output beams ($b_1 = -b_2$). The condition of $1 - M_{12} = 0$ can also be derived similarly, which leads to another set of CPA solutions for the out-of-phase excitation ($a_1 = -a_2$) and lasing solutions with two divergent in-phase output beams ($b_1 = b_2$) as long as the two input beams are not out-of-phase ($a_1 \neq -a_2$). These set of CPA and laser modes appear at a distance $\Delta(\eta) = \pi$ away from the corresponding modes derived earlier from the condition of $1 + M_{12} = 0$. Thus, both CPA and lasing modes repeat themselves with a period of $\Delta(\eta) = \pi$ when the sample thickness is increased. However, their mode structures vary alternately. In other words, the phases of the two input beams for CPA as well as the phases of the two output beams for the laser will switch between in-phase and out-of-phase alternatively.

When $\beta > \alpha > 0$, the CPA mode occurs in the region of $\frac{\pi}{2} < \eta < \pi$ and the lasing mode occurs in the region of $0 < \eta < \frac{\pi}{2}$ instead. Their solutions can be obtained from those of the case of $\alpha > \beta > 0$ by switching between CPA and laser modes.

For the case of $\alpha \neq \beta$ discussed so far, CPA and laser cannot occur simultaneously in a single slab. A numerical demonstration is shown in Fig. 1(c) for the case of $\varepsilon = 1.2i$ and $\mu = -0.5i$. In Fig. 1(c), we fix the frequency at $f$=100THz and calculate the two eigenvalues of the **S**-matrix, $\lambda_\pm$, as a function of $\eta$. The absolute value of the smaller eigenvalue is plotted by a black line and the inverse of the larger eigenvalue by a blue line. The CPA (lasing) is achieved when the black (blue) line touches the $\eta$ axis. Two CPA modes with $\eta$=0.367$\pi$, 1.367$\pi$ and two lasing modes with $\eta$=0.632$\pi$, 1.632$\pi$ are shown in Fig. 1(c). The first and second CPA modes correspond to the in-phase and out-of-phase input beams, respectively.

However, when $\alpha = \beta$, interesting thing happens. From Eqs. (11) and (12), the conditions of CPA and lasing modes obtained from $1+M_{12}=0$ both become $\eta = \pi/2$ ($\sin(\eta)=1$ and $\cos(\eta)=0$) within the region of $0<\eta<\pi$. Interestingly, from Eq. (13), it can be seen that when $\alpha = \beta$, we have $t+r=0$ even although $M_{22}=0$, indicating the realization of CPA. Therefore, CPA and lasing modes can be simultaneously realized, i.e. the so-called "CPA laser".

The coexistence of CPA and laser can also be easily seen from the determinant of the S-matrix, i.e.,

$$|\det(\mathbf{S})|=|\lambda_+\lambda_-|=\left|\frac{2in\mu\cos\eta-(n^2+\mu^2)\sin\eta}{2in\mu\cos\eta+(n^2+\mu^2)\sin\eta}\right|. \quad (14)$$

When $\alpha = \beta$, we have $n^2+\mu^2=0$ and $|\det(\mathbf{S})|=1$. CPA-laser occurs when $\cos(\eta)=0$ at which both numerator and denominator of **S** becomes zero, giving rise to CPA and laser simultaneously. When $\alpha \neq \beta$, $|\det(\mathbf{S})|$ has different expressions for the numerator and denominator and the zeros (CPA) and poles (lasing) occur at different values of . As a demonstration of this case, in Figs. 1 (d), we plot only the absolute value of the smaller eigenvalue of the **S**-matrix as a function of sample thickness for the case of $\varepsilon = 1.2i, \mu = -1.2i$. The inverse of the larger eigenvalue follows the same curve because $|\lambda_+\lambda_-|$=1. The overlap of poles and zeros in this case implies that CPA and laser can be simultaneously achieved in a single slab. However, at each CPA laser solution, the phase relation between two input CPA beams will switch alternately from in-phase to out-of-phase or vice versa when the next solution is reached. At the same time, the phase relation between two output laser beams will switch from out-of phase to in-phase or vice versa.

Finally, we consider the general case of complex $\varepsilon$ and $\mu$ with a real refractive index, e.g., $\varepsilon = 0.6+1.0392i$ and $\mu = 0.25-0.433i$. Here the refractive index is a real number $n=0.7746$. The results in Fig. 1(e) show that neither CPA nor laser can occur at any sample thickness. This can also be understood from Eq. (14). For a complex medium, both the numerator and denominator of det(**S**) are complex functions of frequency and they do not vanish for any real frequency.

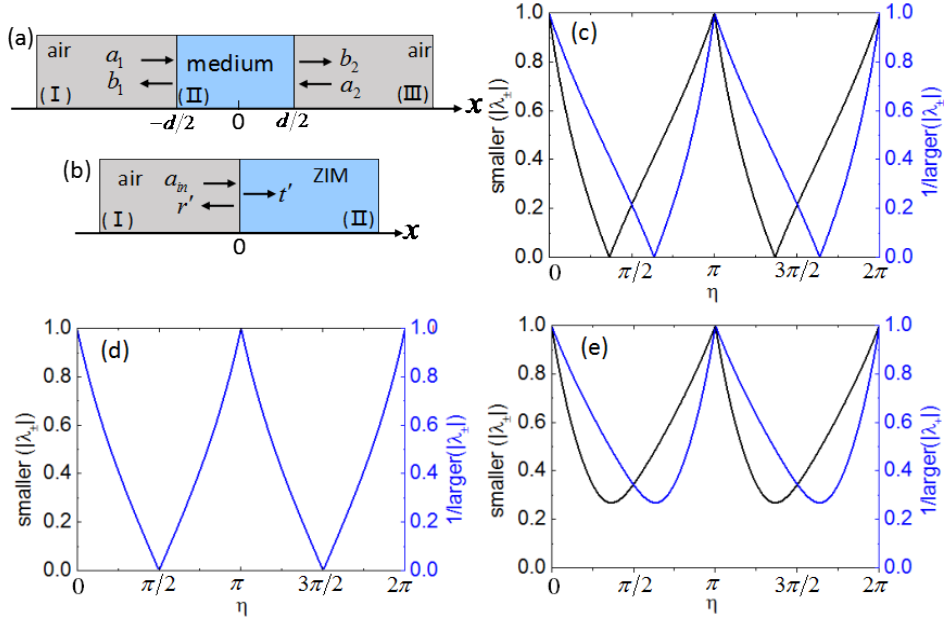

Fig.1 (a) Schematic graph of the ZIM-GL which is symmetric about $x=0$. (b) Schematic graph of the transmission and reflection of plane waves propagating from background to ZIM-GL under normal incidence. (c-e) Values of smaller $(|\lambda_\pm|)$ and $1/\text{larger}(|\lambda_\pm|)$ which are used to determine the poles and zeros of $S$ matrix are plotted as functions of $\eta$. The frequency of the incident wave is $f=100THz$, the complex permittivity and permeability of the media are set as (c): $\varepsilon=1.2i, \mu=-0.5i$; (d): $\varepsilon=1.2i, \mu=-1.2i$; (e): $\varepsilon=0.6+1.0392i, \mu=0.25-0.433i$, respectively.

Now, we study the behaviors and properties of various wave functions in ZIM-GL, particularly for CPA-laser. We should first point out that the lasing mechanism in our system is very different from that of ordinary laser systems. To see this, we consider a plane wave normally incident on a semi-infinite ZIM-GL from one side, as illustrated in Fig. 1(b). The electric and magnetic fields in region I can be written as

$$\begin{cases} E_I = e^{ik_0 x} + r' e^{-ik_0 x} \\ H_I = e^{ik_0 x} - r' e^{-ik_0 x} \end{cases}. \tag{15}$$

Here, $r'$ is the reflection coefficient. Here we have assumed that the polarizations of electric and magnetic fields are along the $y$ and $z$ directions, respectively. The electric and magnetic fields in the ZIM-GL are obtained as

$$\begin{cases} E_{II} = t'e^{ikx} \\ H_{II} = t' \cdot \dfrac{k}{k_0\mu} e^{ikx} \end{cases}, \tag{16}$$

where, $t'$ is the transmission coefficient. By matching the boundary conditions, we obtain

$$\begin{cases} 1+r' = t' \\ 1-r' = -it' \dfrac{1}{\xi} \end{cases}, \tag{17}$$

where $\xi = \sqrt{\dfrac{|\mu|}{|\varepsilon|}}$. We further find that the reflection coefficient can be simplified as

$$r' = \dfrac{-1+i\xi}{1+i\xi}. \tag{18}$$

An interesting conclusion drawn from Eq. (18) is that $|r'|=1$, indicating total reflection of the waves. This is different from ordinary lasers systems, which always have some nonzero transmission, i.e., $|t'| \neq 0$, as the amplified light in the laser cavity cannot penetrate through a hard wall. The absence of energy flux in ZIM-GL can also be seen from the vanishing time-averaged Poynting vector, i.e.,

$$\langle S(x) \rangle_t = \dfrac{1}{2} \mathrm{Re}\left( E_{II} H_{II}^* \hat{y} \times \hat{z} \right)_x = \mathrm{Re}\left( \dfrac{i}{2\xi} |t'|^2 \right) = 0. \tag{19}$$

due to $\pi/2$ phase difference between the electric and magnetic fields as shown in Eq. (16).

Now we show the field patterns of CPA modes. For the case of two in-phase input beams, i.e., $nk_0 d = \dfrac{\pi}{2} + 2M\pi$, the normalized electric and magnetic fields in each region have the following forms

$$\begin{cases} E_I = e^{ik_0 x}, H_I = e^{ik_0 x} \\ E_{II} = \sqrt{2}\cos(nk_0 x), H_{II} = -\sqrt{2}\sin(nk_0 x), \\ E_{III} = e^{-ik_0 x}, H_{III} = -e^{-ik_0 x} \end{cases} \tag{20}$$

The corresponding the time-averaged Poynting vector has the following pattern:

$$\langle S(x) \rangle_t = \begin{cases} \dfrac{1}{2} & x \in \text{I} \\ -\dfrac{1}{2}\sin(2nk_0 x) & x \in \text{II} \\ -\dfrac{1}{2} & x \in \text{III} \end{cases} \qquad (21)$$

A typical curve of Eq. (21) is shown in Fig. 2(a) for $M = 0$. It is clearly shown that the energy flowing in from the background to ZIM-GL is totally absorbed.

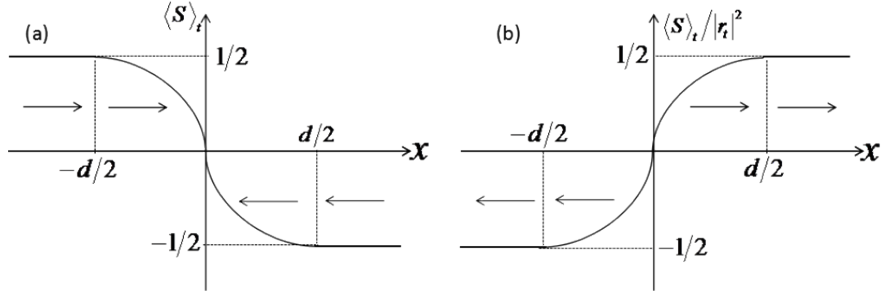

Fig. 2 The time-averaged Poynting vector flow analysis for ZIM-GL in one dimension structure (see Fig. 1(a)) (a) in CPA mode and (b) in lasing mode.

For the laser mode, the electric and magnetic fields in each region are

$$\begin{cases} E_\text{I} = r_t e^{-ik_0 x}, H_\text{I} = -r_t e^{-ik_0 x} \\ E_\text{II} = -\sqrt{2}r_t \sin(nk_0 x), H_\text{II} = -\sqrt{2}r_t \cos(nk_0 x), \\ E_\text{III} = -r_t e^{ik_0 x}, H_\text{III} = -r_t e^{ik_0 x} \end{cases} \qquad (22)$$

where $r_t$ denotes the amplitude of the output beams. Since $r_t$ diverges at the lasing mode, we have ignored the source term in Eq. (22). The corresponding time-averaged Poynting vector now becomes

$$\langle S(x) \rangle_t = \begin{cases} -\dfrac{1}{2}|r_t|^2 & x \in \text{I} \\ \dfrac{1}{2}\sin(2nk_0 x)|r_t|^2 & x \in \text{II} \\ \dfrac{1}{2}|r_t|^2 & x \in \text{III} \end{cases} \qquad (23)$$

A typical curve for the normalized Poynting vector $\langle S(x) \rangle_t / |r_t|^2$ for $M = 0$ is shown in Fig. 2(b). It is interesting to show that Eqs. (20-23) satisfy the following steady state Poynting theorem for harmonic fields:

$$\frac{d}{dx}\langle S(x)\rangle_t = -\frac{k_0}{2}\left[Im\varepsilon |E|^2 + Im\mu |H|^2\right]. \tag{24}$$

In the case of CPA, by taking $Im\varepsilon = -Im\mu = \alpha = n$ and using Eqs. (20) and (21) in Eq. (24), it is easy to show that both sides of Eq. (24) give $\alpha k_0 \cos(2nk_0)$. The similar case is obtained for the laser, i.e., Eqs. (22) and (23) give $\alpha k_0 \cos(2nk_0)|r_t|^2$ on both sides of Eq. (24). Since there is no energy transport between ZIM-GL and its air background, the generation of the Poynting vector in ZIM-GL is solely due to the uneven spatial distributions of the electric and magnetic fields inside ZIM-GL, which in turn acts as a net source or sink of the Poynting vector at every point. This is also very different from ordingary lasers. At CPA-laser, the behaviors of $\langle S(x)\rangle_t$ are identical for CPA and laser except opposite in signs.

Now we show the field patterns for the other set of CPA laser modes at $nk_0 d = \frac{3\pi}{2} + 2M\pi$. The electric and magnetic fields now become

$$\begin{cases} E_I = e^{ik_0 x}, H_I = e^{ik_0 x} \\ E_{II} = \sqrt{2}\sin(nk_0 x), H_{II} = -\sqrt{2}\cos(nk_0 x), \\ E_{III} = -e^{-ik_0 x}, H_{III} = e^{-ik_0 x} \end{cases} \tag{25}$$

and

$$\begin{cases} E_I = r_t e^{-ik_0 x}, H_I = -r_t e^{-ik_0 x} \\ E_{II} = \sqrt{2}r_t \cos(nk_0 x), H_{II} = \sqrt{2}r_t \sin(nk_0 x), \\ E_{III} = r_t e^{ik_0 x}, H_{III} = r_t e^{ik_0 x} \end{cases} \tag{26}$$

respectively. Although the field patterns of the two CPA laser modes are different, their behaviors of the Poynting vector are unchanged and are described by Fig. 2.

### III. DESIGN OF ZIM-GL

The theoretical analysis in Section II shows that a slab of ZIM-GL with the same magnitude for $\varepsilon$ and $\mu$ can realize both CPA and laser simultaneously. However, no naturally-occurring material can satisfy such stringent requirement on the material parameters. In this section, we propose a design map, guided by an effective medium approximation that has been generally applied to metamaterials and photonic crystals [25-30], to realize a ZIM-GL by using a photonic core-shell structure. A schematic is shown in the inset of Fig. 3, a lossy core material with radius $r_1$ is coated by a layer of gain material with radius $r_2$ and arranged in a square lattice structure. The permittivities of the core and shell are denoted by the complex dielectric constants $\varepsilon_1$ and $\varepsilon_2$, respectively. In order to obtain the appropriate parameters of $\varepsilon_1$, $\varepsilon_2$ as well as the corresponding radii $r_1$, $r_2$ for the effective ZIM-GL, we apply a systematic *inverse design* based on an effective medium theory [31]. The details can be found in the Supplementary Materials [24].

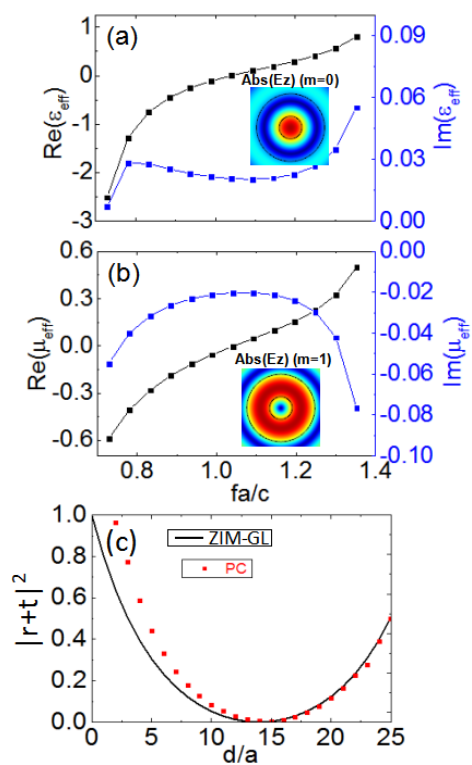

Fig. 3. (a) (b) The effective permittivity $\varepsilon_{eff}$ and permeability $\mu_{eff}$ of the designed photonic structure as functions of frequency. The core-shell structure with inner and outer radii of $r_1 = 0.16a, r_2 = 0.48a$ is designed, where $a$ is lattice constant of the square lattice. The permittivities of the core and shell are set to be $\varepsilon_1 = 1.133 + 0.11i, \varepsilon_2 = 1.07 - 0.054i$, respectively. The dashed lines indicated the CPA-laser point of $fa/c = 1.04$, $\varepsilon_{eff} = 0.02i$,

$\mu_{eff} = -0.02i$. The insets in (a) (b) show the field distributions of the electric fields (absolute value) under the excitation of monopolar and dipolar waves, respectively. (c) The analytical outgoing spectra $|r+t|^2$ as a function of the varying thicknesses $d$ of the ZIM with gain and loss (Black line), and the designed photonic crystal structure (Red square).

We set the working frequency as $f = 100THz$, at which the medium exhibits effective parameters of $\varepsilon_{eff} = 0.02i$ and $\mu_{eff} = -0.02i$. According to the previous analysis, such a medium with the thickness $d = 43.673\mu m$ can act as a CPA-laser, i.e. the simultaneous realization of CPA and laser. The lattice constant is obtained by assuming there are 14 periods along the propagation direction, i.e., $a = d/14 = 3.1195\mu m$, and the inner and outer radii of the core-shell cylinder are set to be $r_1 = 0.16a$, $r_2 = 0.48a$, respectively. Based on the effective medium theory, we can obtain the complex permittivities for the core and shell of cylinders: $\varepsilon_1 = 1.33 + 0.11i$ and $\varepsilon_2 = 1.07 - 0.054i$ (see the Supplementary Material for details). In Fig. 3 we plot the real and imaginary parts of permittivity and permeability in a frequency range according to the following effective medium formulas (see Ref. [31]):

$$\frac{\varepsilon_{eff} + 2\varepsilon_0 \frac{J_0'(k_0 r_0)}{k_0 r_0 J_0(k_0 r_0)}}{\varepsilon_{eff} + 2\varepsilon_0 \frac{Y_0'(k_0 r_0)}{k_0 r_0 Y_0(k_0 r_0)}} = \frac{Y_0(k_0 r_0)}{iJ_0(k_0 r_0)} \left( \frac{D_0}{1+D_0} \right), \quad (27)$$

$$\frac{\mu_{eff} - \mu_0 \frac{J_1(k_0 r_0)}{k_0 r_0 J_1'(k_0 r_0)}}{\mu_{eff} - \mu_0 \frac{Y_1(k_0 r_0)}{k_0 r_0 Y_1'(k_0 r_0)}} = \frac{Y_1'(k_0 r_0)}{iJ_1'(k_0 r_0)} \left( \frac{D_1}{1+D_1} \right), \quad (28)$$

where $J_m(x)$ and $Y_m(x)$ $(m=0,1)$ are the Bessel function of the first and second kind, respectively. $D_m$ represent the Mie scattering coefficients of the core-shell cylinder. As shown in Figs. 3(a) and 3(b), the real parts of $\varepsilon_{eff}$ and $\mu_{eff}$ vanishes simultaneously at frequency $fa/c = 1.04$, where their imaginary parts are close to $\text{Im}(\varepsilon_{eff}) = 0.02i$, and $\text{Im}(\mu_{eff}) = -0.02i$. Such a combination of effective medium parameters satisfies the condition of CPA-laser. The field distributions (absolute value of $E_z$) of the core-shell cylinder under the excitation of monopolar and dipolar waves are displayed in the insets of Figs. 3(a) and 3(b), with dark red represent the maximum value and dark blue represent zero. They correspond to the eigenstates of the structure. It is clearly seen that under the monopolar excitation $(m=0)$, the fields are concentrated in the lossy core, and therefore the majority of the incident wave energy is absorbed by the core; while under the dipolar excitation, the fields are concentrated in the shell with gain medium, thus implying energy enhancement by the gain medium in the shell. In order to further improve the accuracy of the effective parameters beyond the limitation of the effective medium theory, we have performed band structure calculations based on a finite element software, COMSOL Multi-physics, and plot the results in Supplementary Material [24]. By fine-tuning the parameters of the core and shell, we finally obtain an effective medium which almost exactly satisfies the requirement of CPA-laser at a real frequency. The final values of the parameters are $\varepsilon_1 = 1.154 + 0.11i$ and $\varepsilon_2 = 1.07 - 0.054i$, which give rise to effective permittivity and permeability of $\varepsilon_{eff} = 0.01838i$ and $\mu_{eff} = -0.01838i$ at the real frequency of $fa/c = 0.972$. In the next section, we will conduct numerical simulations to verify the functionality of the designed photonic structures as the required effective ZIM-GL.

## IV. NUMERICAL SIMULATIONS OF "CPA LASER"

The scattering property of a slab of the core-shell cylinder structures proposed in Section III is simulated by COMSOL. Figure 3(c) shows the outgoing spectra $|r+t|^2$ with varying thicknesses $d$ of the effective medium of ZIM-GL (Black line) and the real structure (Red square) under the excitation of two counter-propagating coherent waves with the same phase. It is seen that the two sets of results agree with each other, which verifies the functionality of the designed structure as an effective ZIM-GL in terms of transmission properties. From Fig. 3(c), we found that when the thickness of the slab is around $14a$, the outgoing waves vanish, indicating CPA.

The effects of the CPA and laser by using the real photonic core-shell structure is demonstrated in Figs. 4(c) and 5(c), which show the field distributions of electromagnetic wave $E_z$, when two coherent counter-propagating incoming waves are normally incident on to the structure with phase difference $\Delta\varphi=0$ (CPA) and $\Delta\varphi=\pi$ (laser), respectively. The corresponding results obtained from the effective medium of ZIM-GL are shown in Figs. 4(a) and 5(a). In the case of $\Delta\varphi=0$, nearly all the incident waves are absorbed by such a slab of ZIM-GL, as indicated by the total power flow (denoted by the arrows) in Figs. 4(a) and 4(c). In Figs. 4(b) and 4(d), we also plot the amplitudes (normalized to the incident wave) of the electric field for the cases corresponding to Figs. 4(a) and 4(c), respectively. It is clearly seen that for both cases, the electric field in the background is nearly unity, meaning there is no reflection and that almost all of the incident energy is absorbed.

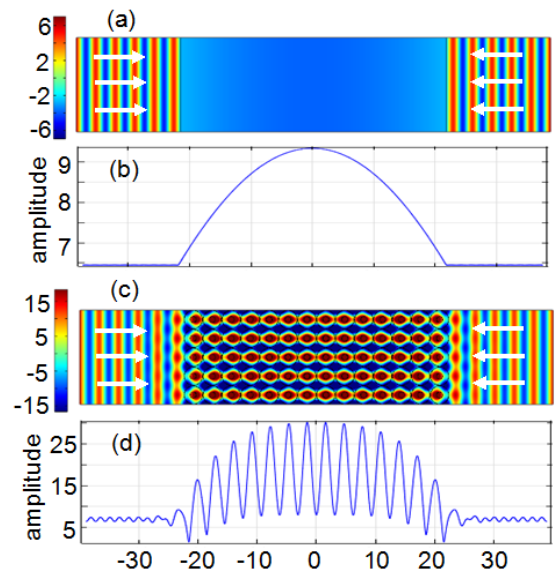

Fig. 4 (a) and (c) show the field distributions of electric field $E_z$ in a slab of the effective medium and the real structure of the ZIM-GL in the case of $\Delta\varphi=0$, respectively. The arrows represent the power flow in the background. (b) and (d) shows the amplitude of electric field $|E_z|$ for the cases of (a) and (c), respectively.

In the case of $\Delta\varphi=\pi$, however, the power flows are completely reversed, as illustrated in Fig. 5(a) and 5(c). The normalized amplitudes of the electric field in the background shown in Figs. 5(b) and 5(d) are found to be about $10^4$ (in effective medium of ZIM-GL) and $10^2$ (in real structure of ZIM-GL) times of the incident waves, indicating a lasing phenomenon. On the other hand, in order to further confirm the lasing mode in photonic structure, we put a row of point sources with line current at the edge of the photonic crystal. The structure is displayed in the insets of Fig. 5(e). The transmission is significantly amplified in the lasing mode and verifies energy enhancement. Fig. 5(e) show the transmission as a function of frequency. A narrow peak is observed in the frequency spectrum with $\Delta(fa/c)=0.021$, at the frequency $fa/c=0.972$ for the ideal lasing mode. Through Figs. 4 and 5, the functionality of CPA-laser is verified.

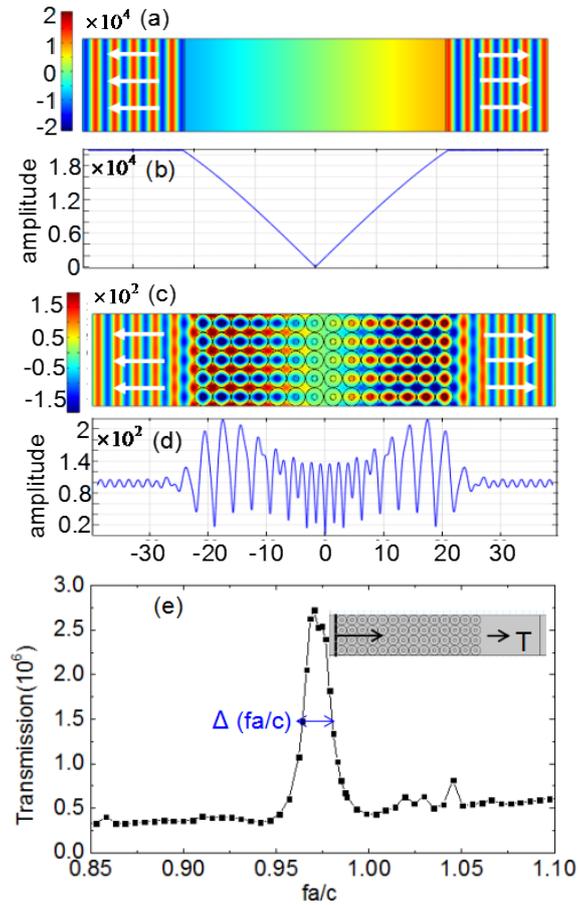

Fig. 5 (a) and (c) show the field distributions of electric field $E_z$ in a slab of the effective medium and the real structures of the ZIM-GL in the case of $\Delta\varphi=\pi$, respectively. The arrows represent the power flow in the background. (b) and (d) shows the amplitude of electric field $|E_z|$ for the cases of (a) and (c), respectively. (e) Transmission as a function of the frequency. The inset in (e) shows the schematic graph of the model.

## V. CONCLUSIONS

In conclusion, we introduce a unique concept of ZIM with its permittivity and permeability being purely imaginary, but exhibiting opposite signs. We demonstrated, through rigorous derivations, that such a ZIM can achieve CPA and laser effect. Especially, when the loss and gain have the same magnitude, the so-called CPA-laser, i.e. simultaneous realization of CPA and laser which was previously only observed in PT-symmetric systems, can be realized in our system without PT symmetry. The functionality of CPA and laser is controlled by the way of excitation, e.g. the symmetry of incident waves hereby.

We also propose a practical scheme to design a ZIM-GL by using photonic structures. By employing the effective medium theory in an inverse manner, and we have successfully designed a square lattice of core-shell dielectric cylinders whose effective medium parameters meet the requirements of the CPA-laser. Numerical simulations verifies the fascinating functionality of switching between a coherent perfect absorber and a laser by altering the symmetry of the excitations, which is consistent with analytical results. Our work demonstrates intriguing possibilities in non-Hermitian optics with both gain and loss distributed in different parameters.

## ACKNOWLEDGMENTS

This work was supported by the State Key Program for Basic Research of China (No. 2014CB360505, No. 2012CB921501), National Natural Science Foundation of China (No. 11374224, 61671314), King Abdullah University of Science and Technology, a Project Funded by the Priority Academic Program Development of Jiangsu Higher Education Institutions (PAPD) and the Area of Excellent Scheme (grant no. AoE/P-02/12).